\documentclass[14pt, fleqn]{extarticle}
\usepackage{latexsym}
\usepackage{amsmath}

\usepackage{amsfonts}
\usepackage{amssymb}
\usepackage{graphicx}
\usepackage{slashbox}
\usepackage{color}
\usepackage{tikz}
\usepgflibrary{arrows}
\usepackage{mathbbol}
\usepackage{xfrac}
\usepackage[a4paper,bindingoffset=0.2in,%
            left=1in,right=1in,top=1in,bottom=1in]{geometry}
\let\OLDthebibliography\thebibliography
\renewcommand\thebibliography[1]{
  \OLDthebibliography{#1}
  \setlength{\parskip}{0pt}
  \setlength{\itemsep}{0pt}
}
\newcommand{\Cross}{\mathbin{\tikz [x=1.4ex,y=1.4ex,line width=.1ex] \draw (0,0) -- (1,1) (0,1) -- (1,0);}}%

\title{Spinning particles, coadjoint orbits\\ and Hamiltonian formalism}
\author{
Krzysztof Andrzejewski\footnote {krzysztof.andrzejewski@uni.lodz.pl}, Cezary Gonera\footnote{cezary.gonera@uni.lodz.pl}, Joanna Gonera\footnote{joanna.gonera@uni.lodz.pl},\\ Piotr Kosi\'nski\footnote {piotr.kosinski@uni.lodz.pl}, Pawe\l{} Ma\'slanka\footnote{pawel.maslanka@uni.lodz.pl}\\
\small \textit {Faculty of Physics and Applied Informatics}\\
\small \textit {University of Lodz, Lodz, Poland}\\
}

\date{}
\begin{document}
\maketitle
\begin{abstract}
\par The extensive analysis of the dynamics of relativistic spinning particles is presented. Using the coadjoint orbits method the Hamiltonian dynamics is explicitly described. The main technical tool is the factorization of general Lorentz transformation into pure boost and rotation. The equivalent constrained dynamics on Poincare group (viewed as configuration space) is derived and complete classification of constraints is performed. It is shown that the first class constraints generate local symmetry corresponding to the stability subgroup of some point on coadjoint orbit. The Dirac brackets for second class constraints are computed. Finally, canonical quantization is performed leading to infinitesimal form of irreducible representations of Poincare group. 
\end{abstract}

\section{Introduction} 
\label{I}
\par The notion of relativistic spinning particles has attracted much attention; a considerable number of papers has been devoted to their description, both on classical and quantum levels. The relevant theories are based on two pillars: the idea of spin as internal angular momentum and the relativistic invariance, i.e. Poincare symmetry.
\par Spin is a peer of modern quantum theory. In attempt to understand its origin various classical models have been proposed \cite{b1}-\cite{b9}; for review see \cite{b10}-\cite{b12}. The underlying ideas have been extended and further developed in a number of more recent papers \cite{b13}, \cite{b14}. Besides these models utilizing commuting variables a separate class of models also exists which uses Grassman variables to describe spin degrees of freedom \cite{b15}-\cite{b19}.
\par Instead of constructing (semi)phenomenological models of spinning particle one can start with the idea of Poincare invariance as the basic principle. Then by the classical model of spinning particle we mean the Hamiltonian dynamical system with Poincare group acting through canonical symmetry transformations on the relevant phase space. We assume also that the action of Poincare group is transitive; this can be viewed as the classical counterpart of irreducibility condition in quantum theory. The emerging possibilities are classified with the help of the orbit method \cite{b20}-\cite{b26} which provides the efficient tool for constructing the relevant Hamiltonian systems. A number of authors follow this approach in describing, both on classical and quantum levels, various dynamical systems exhibiting symmetries \cite{b19}, \cite{b27}-\cite{b30}. In particular, the coadjoint orbits method has been discussed in more detail in \cite{b23}, \cite{b24}, \cite{b31}-\cite{b33} (see also a nice paper \cite{b34} where similar methods are used in the study of geometrical spinoptics). Related approach, utilizing the dynamics defined directly on Poincare group, has been proposed in \cite{b35}-\cite{b40}.
\par In the present paper we provide the complete description of spinning particle within the framework of coadjoint orbit method. Our starting point is the explicit decomposition of general Lorentz transformation into pure boost and rotation \cite{b41}. This allows us to introduce a convenient parametrization of coadjoint orbit describing the massive particles (similar construction for massless particles has been given in \cite{b42}). The resulting Hamiltonian system is described explicitly in terms of this parametrization. Then we construct the Poincare invariant dynamics on Poincare group as configuration space. As compared with direct approach based on coadjoint orbits it contains some redundant variables resulting in additional gauge symmetry based on stability subgroup of canonical point on coadjoint orbit. We present the complete analysis of the resulting constraints of first and second class. We show also how the resulting formalism leads to Wiegmann \cite{b19} parametrization (see also \cite{b33}). Finally, the explicit quantization procedure leading to standard form of Poincare generators is described.
\par In the forthcoming paper the relation of our formalism to the theory of relativistic symmetric top \cite{b35} will be studied.
\par The paper is organized as follows. In Sec. II some basic facts concerning Lorentz and Poincare groups are reminded, mainly in order to fix the notation. Sec. III is devoted to the description of coadjoint orbits of Poincare group corresponding to massive particles. The canonical (Darboux) variables on coadjoint orbit are introduced and the resulting Hamiltonian dynamics is described in detail. In Sec. IV the invariant dynamics on Poincare group is constructed and analysed. Sec. V is devoted to the detailed analysis of the Hamiltonian form of dynamics introduced in Sec. IV. The structure of constraints is exhibited and the relevant Dirac brackets computed. The gauge symmetry generated by two first class constraints is shown to be related to the stability subgroup of some point on coadjoint orbit. In Sec. VI some generalization of action functional is described. It is also shown that the elegant description based on Pauli-Lubanski fourvector results immediately from the formalism discussed in the paper. Sec. VII is devoted to the quantization procedure leading to the unitary irreducible representations of Poincare group. Finally, Sec. VIII contains short summary. In the Appendix we remind the Wigner construction of representations of Poincare group which is the global counterpart of the structure derived in Sec. VII.

\section{Lorentz and Poincare groups} 
\label{II}

\par We remind here some basic facts concerning Lorentz and Poincare groups. Let $ \eta _{\mu \nu} $ be the metric tensor (we adopt the convention $ \eta _{\mu \nu}=\text {diag}(+---) $); Lorentz group consists of four by four matrices $ \Lambda $ obeying
\begin {align} 
\label {al1}
\eta_{\mu \nu}\Lambda^{\mu}_{\phantom {\mu}\alpha}\Lambda^{\nu}_{\phantom{\nu}\beta}=\eta_{\alpha \beta}\quad\text {;}
\end {align}
we restrict ourselves to the proper Lorentz transformations, $ \text {det}\Lambda = 1 $, $ \Lambda ^{0}_{\phantom {0}0}\geqslant 1 $. Infinitesimally,
\begin {align} 
\label {al2}
\Lambda^{\mu}_{\phantom {\mu}\nu}=\delta^{\mu}_{\nu}+\lambda^{\mu}_{\phantom {\mu}\nu}\quad \text {,}\quad\lambda_{\mu\nu}=-\lambda_{\nu\mu}
\end {align}
\par The generators of Lorentz group are represented by the matrices $ M_{\mu \nu}=-M_{\nu \mu} $ obeying
\begin {align} 
\label {al3}
\Lambda^{\alpha}_{\phantom {\alpha}\beta}=\delta^{\alpha}_{\beta}-\frac{i}{2}\lambda^{\mu \nu} (M_{\mu \nu})^{\alpha}_{\phantom {\alpha}\beta}
\end {align}
By virtue of eqs. (\ref{al2}) and (\ref{al3}) one finds
\begin {align} 
\label {al4}
(M_{\mu \nu})^{\alpha}_{\phantom {\alpha}\beta}=i(\delta^{\alpha}_{\mu}g_{\nu \beta}-\delta^{\alpha}_{\nu}g_{\mu \beta})
\end {align}
Poincare group is obtained by supplementing the Lorentz transformations with translations represented by fourvectors $ a^{\mu} $, $ g \equiv (\Lambda, a)$. The composition law reads
\begin {align} 
\label {al5}
(\Lambda, a)\cdot (\Lambda',a')=(\Lambda \Lambda',\Lambda a'+a)
\end {align}
while the inverse is given by
\begin {align} 
\label {al6}
(\Lambda, a)^{-1}=(\Lambda^{-1},-\Lambda^{-1}a)
\end {align}
\par An infinitesimal element of Poincare group can be written as
\begin {align} 
\label {al7}
g=1+i \epsilon^{\mu}P_{\mu}-\frac{i}{2}\lambda^{\mu \nu}M_{\mu \nu}\quad\text{;}
\end {align}
here $ P_{\mu} $ and $ M_{\mu \nu} $ are understood as abstract elements of Poincare Lie algebra. Eqs. (\ref{al5}) and (\ref{al7}) imply the following commutation rules defining this algebra:
\begin {align} 
\label {al8}
[M_{\mu \nu}, P_{\alpha}]=i(\eta_{\nu \alpha}P_{\mu}-\eta_{\mu \alpha}P_{\nu})
\end {align}
\begin {align} 
\label {al9}
[M_{\mu \nu}, M_{\alpha \beta}]=i(\eta_{\mu \beta}M_{\nu \alpha}+\eta_{\nu \alpha}M_{\mu \beta}-\eta_{\mu \alpha}M_{\nu \beta}-\eta_{\nu \beta}M_{\mu \alpha})
\end {align}
\begin {align} 
\label {al10}
[P_{\mu},P_{\nu}]=0
\end {align}
\par Poincare algebra admits two independent Casimir operators, mass squared
\begin {align} 
\label {al11}
M^{2}\equiv P^{\mu}P_{\mu}
\end {align}
and the square of Pauli-Lubanski fourvector,
\begin {align} 
\label {al12}
W^{2}\equiv W^{\mu}W_{\mu}\quad \text {,} \quad W^{\mu} \equiv \frac{1}{2}\epsilon^{\mu \nu \alpha \beta}P_{\nu}M_{\alpha \beta}
\end {align}
The adjoint action of Poincare group on its Lie algebra reads $\big(g \equiv (\Lambda, a)\big) $:
\begin {align} 
\label {al13}
gP_{\mu}g^{-1}=\Lambda^{\nu}_{\phantom {\nu}\mu }P_{\nu}
\end {align}
\begin {align} 
\label {al14}
gM_{\mu \nu}g^{-1}=\Lambda^{\alpha}_{\phantom {\alpha}\mu }\Lambda^{\beta}_{\phantom {\beta}\nu }M_{\alpha \beta}+(\Lambda^{\alpha}_{\phantom {\alpha}\mu }\Lambda^{\beta}_{\phantom {\beta}\nu }-\Lambda^{\alpha}_{\phantom {\alpha}\nu }\Lambda^{\beta}_{\phantom {\beta}\mu })a_{\beta}P_{\alpha}
\end {align}

\section{Coadjoint orbits and phase space for massive particles} 
\label{III}
\par We shall consider Hamiltonian dynamics invariant under the action of Poincare group. It is assumed that the latter acts transitively on the phase space; such a system may be viewed as elementary. The Hamiltonian dynamics is constructed using the coadjoint orbits method \cite{b20}-\cite{b26}. We start with the space dual to the Poincare algebra. The relevant coordinates, corresponding to the generators $P_{\mu}$ and $M_{\mu \nu}$, will be denoted by $\zeta_{\mu}$ and 
$\zeta_{\mu \nu}=-\zeta_{\nu \mu}$, respectively.
\par We impose the Poisson structure implied by the Lie algebra commutation rules\begin {align} 
\label {al15}
\{\zeta_{\mu},\zeta_{\nu}\}=0
\end {align}
\begin {align} 
\label {al16}
\{\zeta_{\mu \nu},\zeta_{\alpha}\}=\eta_{\nu \alpha}\zeta_{\mu}-\eta_{\mu \alpha}\zeta_{\nu}
\end {align}
\begin {align} 
\label {al17}
\{\zeta_{\mu \nu},\zeta_{\alpha \beta}\}=\eta_{\mu \beta}\zeta_{\nu \alpha}+\eta_{\nu \alpha}\zeta_{\mu \beta}-\eta_{\mu \alpha}\zeta_{\nu \beta}-\eta_{\nu \beta}\zeta_{\mu \alpha}
\end {align}
It is invariant under the coadjoint action of Poincare group which reads
\begin {align} 
\label {al18}
\mathrm{Ad}^{*}_{(\Lambda,a)}\zeta_{\mu}=\Lambda_{\mu}^{\phantom{\mu}\nu}\zeta_{\nu}
\end {align}
\begin {align} 
\label {al19}
\mathrm{Ad}^{*}_{(\Lambda,a)}\zeta_{\mu \nu}=\Lambda_{\mu}^{\phantom{\mu}\alpha}\Lambda_{\nu}^{\phantom{\nu}\beta}\zeta_{\alpha \beta}-(\Lambda_{\mu}^{\phantom{\mu}\alpha}a_{\nu}-\Lambda_{\nu}^{\phantom{\nu}\alpha}a_{\mu})\zeta_{\alpha}
\end {align}
There are two functionally independent invariants which correspond to two Casimir operators and characterize the generic coadjoint orbits
\begin {align} 
\label {al20}
m^{2}=\zeta^{\mu}\zeta_{\mu}
\end {align}
\begin {align} 
\label {al21}
-m^{2}s^{2}=w^{\mu}w_{\mu}\quad \text {,}\quad w^{\mu}\equiv \frac{1}{2}\epsilon^{\mu \nu \alpha \beta}\zeta_{\nu}\zeta_{\alpha \beta}
\end {align}
The components $w_{\mu}$ obey the following Poisson commutation rules:
\begin {align} 
\label {al22}
\{w ^{\mu},\zeta_{\nu}\}=0
\end {align}
\begin {align} 
\label {al23}
\{w ^{\mu},\zeta_{\alpha \beta}\}=\delta^{\mu}_{\alpha}w_{\beta}-\delta^{\mu}_{\beta}w_{\alpha}
\end {align}
\begin {align} 
\label {al24}
\{w ^{\mu},w^{\nu}\}=\epsilon^{\mu \nu \alpha \beta}\zeta_{\alpha}w_{\beta}
\end {align}
In what follows we consider the orbits corresponding to $\zeta^{\mu}\zeta_{\mu}\equiv m^{2}>0$; it is also easy to see that $\zeta_{0}>0$ is an invariant condition which we assume to hold true. Fixing the values of $\zeta^{\mu}\zeta_{\mu}$ and $w^{\mu}w_{\mu}$ we obtain eightdimensional manifold which becomes our phase space; indeed, the key point is that the Poisson structure \cite{b15}-\cite{b17}, when restricted to an orbit, becomes nondegenerate \cite{b20}-\cite{b26}.
\par It is quite easy to find a simple canonical point on any orbit described above. First note that by coadjoint action of Lorentz group one can put $\zeta_{\mu}$ in form
\begin {align} 
\label {al25}
\underline{\zeta}_{\mu}=(m,\vec{0})\quad {;}
\end {align}
$\underline{\zeta}_{\mu}$ is further left unchanged by any rotation. On the other hand, under the rotation subgroup, $\zeta_{ij}$ transforms as a second order antisymmetric tensor. Puting
\begin {align} 
\label {al26}
\zeta _{ij}=s_{k}\epsilon_{kij}
\end {align}
we conclude that $\vec{s}$ is a threevector. By rotation it can be put in the form $\vec{s}=(0,0,s)$ (its magnitude is determined by the equation  (\ref{al21})). Therefore,
\begin {align} 
\label {al27}
\underline{\zeta}_{ij}=s \epsilon_{3ij}
\end {align}
\par It remains to consider $\zeta_{0i}$. Eq. (\ref{al19}) implies the following transformation rule for $\zeta_{0i}$ under $\mathrm{Ad}^{*}_{(1,a)}$
\begin {align} 
\label {al28}
\zeta '_{0i}=\zeta_{0i}-ma_{i}
\end {align}
while leaving $\underline{\zeta}_{\mu}$ and $\underline{\zeta}_{ij}$ invariant. Due to $m \neq 0$ we can put $\zeta _{0i}=0$.\\
Summarizing, the canonical point on the orbit $\zeta^{\mu}\zeta_{\mu}=m^{2}>0$, $\zeta^{0}>0$, can be chosen as
\begin {align} 
\label {al29}
\underline {\zeta}_{\mu}=(m,\vec{0})
\end {align}
\begin {align} 
\label {al30}
\underline {\zeta}_{0i}=-\underline{\zeta}_{i0}=0
\end {align}
\begin {align} 
\label {al31}
\underline {\zeta}_{ij}=s\epsilon_{3ij}
\end {align}
The stability subgroup of the above canonical point is $G_{s}=SO(2)\times \mathbb{R}$ with $SO(2)$ being rotations around third axis while $\mathbb{R}$ is the additive group of time translations. 
\par The next step is the convenient parametrization of the orbit under consideration (which is isomorphic to the coset space Poincare/$G_{s}$). To this end we decompose the general Lorentz transformation into pure boost and rotation (cf. \cite{b41})
\begin {align} 
\label {al32}
\Lambda=L \cdot R
\end {align}
where
\begin {align} 
\label {al33}
L^{\mu}_{\phantom {\mu}0}\equiv L^{0}_{\phantom {0}\mu}=\Lambda^{\mu}_{\phantom {\mu}0}
\end {align}
\begin {align} 
\label {al34}
L^{i}_{\phantom {i}j}\equiv \delta^{i}_{j}-\frac{\Lambda^{i}_{\phantom {i}0}\Lambda_{j}^{\phantom {j}0}}{1 + \Lambda^{0}_{\phantom{0}0}}
\end {align}
\begin {align} 
\label {al35}
R^{\mu}_{\phantom {\mu}0}=R^{0}_{\phantom {0}\mu}=\delta^{\mu}_{0}
\end {align}

\begin {align} 
\label {al36}
R^{i}_{\phantom {i}j}=\Lambda^{i}_{\phantom {i}j}-\frac{\Lambda^{i}_{\phantom {i}0}\Lambda^{0}_{\phantom{0}j}}{1 + \Lambda^{0}_{\phantom{0}0}}
\end {align}
The rotation matrix can be further decomposed as follows
\begin {align} 
\label {al37}
R=\tilde{R}\cdot R_{3}
\end {align}
where $\tilde {R}$ is an element of the coset manifold $SO(3)/SO(2)$ which is the twodimensional sphere $S_{2}$ consisting of directions of $\vec{s}$ while $R_{3}$ is a rotation around third axis. This decomposition is local because $SO(3)$ as the $SO(2)$ bundle over $S_{2}$ is nontrivial (Hopf) bundle. Concluding, any element $\Lambda$ of Lorentz group can be factorized as follows 
\begin {align} 
\label {al38}
\Lambda=L \cdot \tilde{R} \cdot R_{3}
\end {align}
Now, let us find the corresponding factorization of Poincare group. To this end let  $z=(z^{0},\vec{0})$, $y=(0,\vec{y})$; consider any element $(\Lambda,a)$ of Poincare group and write
\begin {align} 
\label {al39}
(\Lambda,a)=(1,y)(L,0)(\tilde{R},0)(R_{3},z)
\end {align}
Using the multiplication rule (\ref{al5}) one finds
\begin {align} 
\label {al40}
z^{0}=\frac{a^{0}}{L^{0}_{\phantom {0}0}}
\end {align}
\begin {align} 
\label {al41}
y^{i}=a^{i}-\frac{L^{i}_{\phantom {i}0}a^{0}}{L^{0}_{\phantom {0}0}}
\end {align}
Eqs. (\ref{al39})-(\ref{al41}) define the parametrization of the Poincare group decomposed into the cosets with respect to the stability subgroup $G_{s}$. 
\par The coadjoint action of Poincare group on the canonical point $(\underline{\zeta}_{\mu},\underline{\zeta}_{\mu \nu})$ generates the relevant orbit. Therefore,
\begin {align} 
\label {al42}
(\zeta_{\mu},\zeta_{\mu \nu})=(\Lambda,a)\ast (\underline{\zeta}_{\mu},\underline{\zeta}_{\mu \nu})=\big( (1,y)(L,0)(\tilde{R},0)\big)\ast (\underline{\zeta}_{\mu},\underline{\zeta}_{\mu \nu})
\end {align}
where star denotes coadjoint action. Using eqs. (\ref{al18}) and (\ref{al19}) and taking into account that $\tilde{R}$ rotates spin to its actual direction we find
\begin {align} 
\label {al43}
\zeta_{\mu}=m\Lambda_{\mu}^{\phantom {\mu}0}\equiv p_{\mu}
\end {align}
\begin {align} 
\label {al44}
\zeta_{0i}=-p_{0}y_{i}+\frac{\epsilon_{ijk}s_{j}p_{k}}{m}
\end {align}
\begin {align} 
\label {al45}
\zeta_{ij}=-p_{i}y_{j}+p_{j}y_{i}+s_{k}\epsilon_{kij}+\frac{(p_{i}\epsilon_{jlk}-p_{j}\epsilon_{ilk})s_{l}p_{k}}{m(m+p^{0})}
\end {align}
Eq. (\ref{al45}) is slightly complicated. However, by redefining
\begin {align} 
\label {al46}
x_{i}\equiv y_{i}-\frac{\epsilon_{ilk}s_{l}p_{k}}{m(m+p_{0})}
\end {align}
eqs. (\ref{al44}), (\ref{al45}) take the form
\begin {align} 
\label {al47}
\zeta_{0i}=-p_{0}x_{i}+\frac{\epsilon_{ilk}s_{l}p_{k}}{m+p_{0}}
\end {align}
\begin {align} 
\label {al48}
\zeta_{ij}=x_{i}p_{j}-x_{j}p_{i}+s_{k}\epsilon_{kij}
\end {align}
We have defined the parametrization of coadjoint orbit in terms of $x_{i}$, $p_{i}$ and $s_{i}$ (obeying $\vec{s}\,^{2}=s^{2})$. This is our eightdimensional phase space. Now, we can find the relevant Poisson brackets using eqs. (\ref{al15})-(\ref{al17}), (\ref{al43}), (\ref{al47}) and (\ref{al48}). They read
\textsc{\begin {align} 
\label {al49}
\{p_{i},p_{j}\}=0
\end {align}}
\begin {align} 
\label {al50}
\{x_{i},p_{j}\}=\delta_{ij}
\end {align}
\begin {align} 
\label {al51}
\{x_{i},x_{j}\}=0
\end {align}
\begin {align} 
\label {al52}
\{x_{i},s_{j}\}=0=\{p_{i},s_{j}\}
\end {align}
\begin {align} 
\label {al53}
\{s_{i},s_{j}\}=\epsilon_{ijk}s_{k}
\end {align}
One can view $x_{i}$, $p_{i}$ and $s_{i}$ as coordinate, momentum and spin components, respectively.\\
Now, $\zeta_{\mu}$ and $\zeta_{\mu \nu}$ are the generators of Poincare transformations at $t=0$. In particular, $\zeta_{0}\equiv p_{0}$ generates time translations, i.e. it is the Hamiltonian of our system. The resulting canonical equations of motion read
\begin {align} 
\label {al54}
\frac{dx_{i}}{dt}=\{x_{i},p_{0}\}=\frac{p_{i}}{p_{0}}
\end {align}
\begin {align} 
\label {al55}
\frac{dp_{i}}{dt}=\{p_{i},p_{0}\}=0
\end {align}
\begin {align} 
\label {al56}
\frac{ds_{i}}{dt}=\{s_{i},p_{0}\}=0
\end {align}
\par The integrals of motion, corresponding to the initial values $\zeta_{\mu}$ and $\zeta_{\mu \nu}$, read
\begin {align} 
\label {al57}
\zeta_{\mu}(t)=\zeta_{\mu}
\end {align}
\begin {align} 
\label {al58}
\zeta_{ij}(t)=\zeta_{ij}
\end {align}
\begin {align} 
\label {al59}
\zeta_{0i}(t)=\zeta_{0i}+t\zeta_{i}
\end {align}
In the Hamiltonian formalism they generate the Poincare transformations at the moment $t$. Consider the infinitesimal Lorentz transformation rule; it reads
\begin {align} 
\label {al60}
\delta(\cdot)=\{\lambda^{\mu \nu}\zeta_{\mu \nu}(t),(\cdot)\}
\end {align}
Eqs. (\ref{al49})-(\ref{al53}) and (\ref{al58}) imply the standard transformation rules under space rotations (i.e. $x_{i}$, $p_{i}$ and $s_{i}$ form threevectors). On the other hand, the Lorentz boosts get modified. Consider the transformation rule for coordinates. Let $\lambda_{0i}\equiv \delta v_{i}$ denote the infinitesimal velocity. By virtue of eq. (\ref{al60}) one finds
\begin {align} 
\label {al61}
\delta x_{i}=-t\delta v_{i}+\delta v_{k} x_{k}\frac{p_{i}}{p_{0}}+\frac{\epsilon_{ilk}s_{l}\delta v_{k}}{m+p_{0}}-\frac{\epsilon_{kln}\delta v_{k}s_{l}p_{n}p_{i}}{(m+p_{0})^{2}p_{0}}
\end {align}
\par The first term on the right hand side corresponds to the usual Lorentz transformation; the second accounts for the change of time variable (one should keep in mind that in the Hamiltonian formalism time variable is kept fixed so we have to recompute everything back to initial time). The remaining two terms represent spin dependent shift of the coordinate. The existence of such a shift in the massive Dirac particles has been noted in \cite{b43}.
\par Similar results concerning the explicit description of coadjoint orbit as the phase space describing relativistic dynamics have been obtained in Ref \cite{b44}\footnote[1]{We are grateful to J. Varilly for bringing this reference to our attention.}. 

\section{Invariant dynamics} 
\label{IV}
\par Alternatively, the dynamics of spinning relativistic particles may be formulated using the manifold of Poincare group as configuration space with Poincare symmetry acting through left (say) multiplication. Such a description involves redundant variables since the coadjoint orbit is only eightdimensional. As a result the relevant dynamics on Poincare group should exhibit local symmetry under the right action of stability subgroup $G_{s}$ of the canonical point on coadjoint orbit.
\par Guided by the above intuitive argument we are looking for the action functional invariant under the global action of Poincare group (acting by the left multiplication) and local action of the stability subgroup $G_{s}$ (acting by the right multiplication). This can be easily achieved by looking at the construction of Kirillov form \cite{b22}-\cite{b26}, \cite{b33}. Consider the left invariant one-form on Poincare group
\begin {align} 
\label {al62}
g^{-1}dg=(\Lambda^{-1}d \Lambda,\Lambda^{-1}da)
\end {align}
By virtue of eq. (\ref{al7}) one can write 
\begin {align} 
\label {al63}
g^{-1}dg=i (\Lambda^{-1}da)^{\mu}P_{\mu}-\frac{i}{2}(\Lambda^{-1}d \Lambda)^{\mu \nu}M_{\mu \nu}
\end {align}
By replacing in the above formula $(P_{\mu}, M_{\mu \nu})$ by the coordinates of the canonical point, $(\underline{\zeta}_{\mu},\underline{\zeta}_{\mu \nu})$, one finds the relevant one-form
\begin {align} 
\label {al64}
\eta = m(\Lambda^{-1}da)^{0}-\frac{s}{2}\epsilon_{3kl}(\Lambda^{-1}d \Lambda)^{kl}
\end {align}
The action functional
\begin {align} 
\label {al65}
S=-\int \eta =-\int \big(m \Lambda_{\mu}^{\phantom {\mu}0}\dot{a}^{\mu}-\frac{s}{2}\epsilon_{3kl}\Lambda_{\mu}^{\phantom {\mu}k}\dot{\Lambda}^{\mu l}\big)d \tau\quad \text {,}
\end {align}
with $\tau$ being some invariant parameter (taken as an evolution parameter), is invariant under global left Poincare group action and local right $G_{s}$ action (up to the boundary terms) \cite{b36}-\cite{b40}, \cite{b33} (dot denotes differentiation with respect to $\tau$). The relevant Lagrangian takes the form:
\begin {align} 
\label {al66}
L=-m \Lambda_{\mu}^{\phantom {\mu}0}\dot{a}^{\mu}+\frac{s}{2}\epsilon_{3kl}\Lambda_{\mu}^{\phantom {\mu}k}\dot{\Lambda}^{\mu l}
\end {align}
Let us compute the variation $\delta S$. To this end let us remind that $\Lambda^{\mu}_{\phantom {\mu}\nu}$ are not independent variables; they are constrained by eq. (\ref{al1}) which leaves six independent variables. Let us define six independent variations as \cite{b35}
\begin {align} 
\label {al67}
\delta I ^{\mu \nu}\equiv \Lambda^{\mu}_{\phantom {\mu}\rho}\delta \Lambda ^{\nu \rho}\quad \text {,} \quad \delta I^{\mu \nu}=-\delta I ^{\nu\mu}
\end {align}
or
\begin {align} 
\label {al68}
\delta \Lambda^{\mu \nu}=\Lambda_{\alpha}^{\phantom {\alpha}\nu}\delta I^{\alpha \mu}
\end {align}
Using eqs. (\ref{al67}), (\ref{al68}) one finds the following expression for $\delta S$:
\begin {align} 
\label {al69}
\delta S = &- \int \limits ^{\tau_{1}}_{\tau_{0}} \bigg[\frac{1}{2}\big (m (\Lambda_{\alpha}^{\phantom {\alpha}0}\dot{a}_{\mu}-\Lambda_{\mu}^{\phantom {\mu}0}\dot{a}_{\alpha})-s \epsilon_{3kl}(\dot{\Lambda}_{\mu}^{\phantom {\mu}l}\Lambda_{\alpha}^{\phantom {\alpha}k}-\dot{\Lambda}_{\alpha}^{\phantom {\alpha}l}
\Lambda_{\mu}^{\phantom {\mu}k})\delta I^{\alpha \mu}\big)\nonumber\\
&-m\dot{\Lambda}_{\mu}^{\phantom {\mu}0}\delta a ^{\mu}\bigg] d \tau- (m \Lambda_{\mu}^{\phantom {\mu}0}\delta a^{\mu}-\frac{s}{2}\epsilon _{3kl}\Lambda_{\mu}^{\phantom {\mu}k}\Lambda_{\alpha}^{\phantom {\alpha}l}\delta I^{\alpha \mu}\big )\Bigg \vert^{\tau_{1}}_{\tau_{0}}
\end {align} 
Assuming $\delta a_{\mu}(\tau_{1,2})=0$, $\delta I^{\alpha \mu}(\tau _{1,2})=0$ one obtains the equations of motion:
\begin {align} 
\label {al70}
m \dot{\Lambda}_{\mu}^{\phantom {\mu}0}=0
\end {align}
\begin {align} 
\label {al71}
m \Lambda_{\alpha}^{\phantom {\alpha}0}\dot{a}_{\mu}-m \Lambda_{\mu}^{\phantom {\mu}0}\dot{a}_{\alpha}- s\epsilon_{3kl}(\Lambda_{\alpha}^{\phantom {\alpha}k}\dot{\Lambda}_{\mu}^{\phantom {\mu}l}+\dot{\Lambda}_{\alpha}^{\phantom {\alpha}k}\Lambda_{\mu}^{\phantom {\mu}l})=0
\end {align}
Eqs. (\ref{al70}), (\ref{al71}) imply the following conservation laws
\begin {align} 
\label {al72}
\frac{d}{d\tau}(m \Lambda _{\mu}^{\phantom {\mu}0})=0
\end {align}
\begin {align} 
\label {al73}
\frac{d}{d\tau}(m \Lambda _{\alpha}^{\phantom {\alpha}0}a_{\mu}-m\Lambda _{\mu}^{\phantom {\mu}0}a_{\alpha}-s \epsilon _{3kl}\Lambda _{\alpha}^{\phantom {\alpha}k}\Lambda _{\mu}^{\phantom {\mu}l})=0
\end {align}
Eqs. (\ref{al70}), (\ref{al71}) may be simplified as follows. Multiplying eq. (\ref{al71}) by $\Lambda^{\mu 0}$ and using (\ref{al70}) one finds
\begin {align} 
\label {al74}
m\frac{da_{\alpha}}{d\tau}=m(\dot{a}_{\mu}\Lambda^{\mu}_{\phantom {\mu}0})\Lambda_{\alpha}^{\phantom {\alpha}0}
\end {align}
Inserting this back to eq. (\ref{al71}) yields
\begin {align} 
\label {al75}
\frac{d}{d\tau}(s\epsilon_{3kl}\Lambda_{\mu}^{\phantom {\mu}k}\Lambda_{\nu}^{\phantom{\nu}l})=0
\end {align}
One can view eqs. (\ref{al70}), (\ref{al74}), (\ref{al75}) as the final form of equations of motion. 
\par The conservation laws (\ref{al72}), (\ref{al73}) result from global Poincare symmetry. Indeed, the infinitesimal Poincare transformations read
\begin {align} 
\label {al76}
(\Lambda,a)\rightarrow (1+\lambda, \epsilon)(\Lambda,a)=(\Lambda+\lambda \Lambda,a+\lambda a+ \epsilon)
\end {align}
or, explicitly
\begin {align} 
\label {al77}
&\delta \Lambda^{\mu}_{\phantom{\mu}\nu}=\lambda^{\mu}_{\phantom{\mu}\alpha}\Lambda^{\alpha}_{\phantom{\alpha}\nu}\nonumber\\
&\delta a ^{\mu}=\lambda^{\mu}_{\phantom {\mu}\nu}a^{\nu}+\epsilon^{\mu}\quad \text {,}\quad
\lambda^{\mu \nu}=-\lambda^{\nu \mu}
\end {align}
By virtue of eq. (\ref{al69}) one concludes that $m\Lambda_{\mu}^{\phantom{\mu}0}$ and $m \Lambda_{\alpha}^{\phantom {\alpha}0}a_{\mu}-m\Lambda_{\mu}^{\phantom {\mu}0}a_{\alpha}-s \epsilon_{3kl}\Lambda_{\alpha}^{\phantom {\alpha}k}\Lambda_{\mu}^{\phantom {\mu}l}$ are conserved on-shell, in accordance with eqs. (\ref{al72}), (\ref{al73}).\\
Consider now the right local action of the stability subgroup $G_{s}=SO(2)\times \mathbb{R}$. It reads
\begin {align} 
\label {al78}
(\Lambda,a)(1+\delta R,\epsilon)=(\Lambda+\Lambda \delta R, a+\Lambda \epsilon)
\end {align}
where
\begin {align} 
\label {al79}
\epsilon=(\epsilon^{0}(\tau),\vec{0})
\end {align}
while
\begin {align} 
\label {al80}
\delta R^{\mu}_{\phantom{\mu}\nu}=\lambda(\tau)(\delta^{\mu}_{1} \delta^{2}_{\nu}-\delta^{\mu}_{2}\delta^{1}_{\nu})
\end {align}
is the rotation around third axis by an infinitesimal angle $\lambda (\tau)$. Combining these equations with eq. (\ref{al69}) one concludes that no nontrivial conservation laws are obtained if we put $\epsilon^{0}(\tau)=\epsilon^{0}\equiv \text {const}$, $\lambda(\tau)=\lambda\equiv \text {const}$. This agrees with the general form of second Noether theorem when applied to our particular situation. However, $G_{s}$ is the local symmetry so it implies some Bianchi identities. They read
\begin {align} 
\label {al81}
\Lambda^{\mu}_{\phantom {\mu}0}\frac{d}{d\tau}(m \Lambda_{\mu}^{\phantom {\mu}0})\equiv 0
\end {align}
\begin {align} 
\label {al82}
(\Lambda^{\mu}_{\phantom {\mu}1}\Lambda^{\nu}_{\phantom {\nu}2}-\Lambda^{\mu}_{\phantom {\mu}2}\Lambda^{\nu}_{\phantom {\nu}1})\big (m\Lambda_{\nu}^{\phantom{\nu}0}\dot{a}_{\mu}-m \Lambda_{\mu}^{\phantom{\mu}0}\dot{a}_{\nu}-s\epsilon_{3kl}(\dot{\Lambda}_{\mu}^{\phantom {\mu}l}\Lambda_{\nu}^{\phantom {\nu}k}+\Lambda_{\mu}^{\phantom {\mu}l}\dot{\Lambda}_{\nu}^{\phantom {\nu}k})   \big)\equiv 0
\end {align}
for $\mathbb{R}$ and $SO(2)$ components of $G_{s}$, respectively; both eqs. (\ref{al81}) and (\ref{al82}) are identities valid of-shell.
\par It is not difficult to show that our equations of motion agree with those derived in previous section. The relevant variables defined in Sec. III are $x_{i}$, $p_{i}$ and $s_{i}$. Eqs. (\ref{al43}) and (\ref{al70}) imply
\begin {align} 
\label {al83}
\frac{dp_{\mu}}{d \tau}=0
\end {align}
which is invariant under arbitrary $\tau$ reparametrization yielding (\ref{al55}) if $\tau$ is chosen to be physical time (see also below). \\
Note further that eq. (\ref{al74}) can be rewritten as
\begin {align} 
\label {al84}
\frac{da^{\mu}}{d\tau}=\frac{p^{\mu}p_{\nu}}{m^{2}}\frac{da^{\nu}}{d \tau}
\end {align}
It is invariant under the local translations belonging to the stability subgroup (cf. eq. (\ref{al78})):
\begin {align} 
\label {al85}
p_{\mu}\rightarrow p'_{\mu}=p_{\mu}
\end {align}
\begin {align} 
\label {al86}
a^{\mu}\rightarrow a'^{\mu}=a^{\mu}+\Lambda^{\mu}_{\phantom {\mu}0}\epsilon^{0}(\tau)=a^{\mu}+\frac{p^{\mu}}{m}\epsilon^{0}(\tau)
\end {align}
with $\epsilon^{0}(\tau)$ being an arbitrary function of $\tau$. Choosing the gauge $a^{0}=\tau$ one finds from eq. (\ref{al84})
\begin {align} 
\label {al87}
1=\frac{p^{0}p_{\nu}}{m^{2}}\frac{da^{\nu}}{d a^{0}}
\end {align}
and, consequently
\begin {align} 
\label {al88}
\frac{da_{i}}{da^{0}}=\frac{p_{i}}{p_{0}}
\end {align}
Combining eqs. (\ref{al41}) and (\ref{al46}) we find that
\begin {align} 
\label {al89}
x_{i}=a_{i}-\frac{p_{i}a_{0}}{p_{0}}-\frac{\epsilon_{ilk}s_{l}p_{k}}{m(m+p_{0})}
\end {align}
\par We shall show below that $s_{i}$ are constant along trajectories. By virtue of eq. (\ref{al84}) we find
\begin {align} 
\label {al90}
\frac{dx_{i}}{d\tau}=0
\end {align}
so, identifying $a^{\mu}$ with Minkowski space-time coordinates we find that $x_{i}$ is, basically, the initial value, at $a^{0}=0$, of the coordinate $a_{i}$. Then $x_{i}(t)$ is the initial value of $a_{i}$ at $a^{0}=t$. Therefore, 
\begin {align} 
\label {al91}
x_{i}(t)=a_{i}(a_{0})-\frac{p_{i}}{p_{0}}(a_{0}-t)
\end {align}
which implies eq. (\ref{al54}).
\par We have related the gauge-invariant (with $G_{s}$ as gauge group) combinations of $a^{\mu}$ and $\Lambda^{\mu}_{\phantom {\mu}0}$ to $x_{i}$, $p_{i}$ and $s_{i}$. The gauge invariant combinations of the orthogonal matrix $R$ (cf. eqs. (\ref{al35}), (\ref{al36})) belong to the coset manifold $SO(3)/SO(2)$, i.e. they are expressible in terms of spin variables $s_{i}$ which parametrize the two-sphere $\vec{s}\,^{2}=s^{2}$. In order to find the equation of motion for $s_{i}$ we start with eq. (\ref{al75}). Inserting there the decomposition (\ref{al32}) and keeping in mind that $L^{\mu}_{\phantom{\mu}\nu}$ are constant elements of invertible matrix we find
\begin {align} 
\label {al92}
\frac{d}{d\tau}(s \epsilon_{3kl}R_{m}^{\phantom {m}k}R_{n}^{\phantom {n}l})=0
\end {align}
However, by construction
\begin {align} 
\label {al93}
s \epsilon_{3kl}R_{m}^{\phantom {m}k}R_{n}^{\phantom {n}l}=s \epsilon_{kmn}R_k^{\phantom {k}3}\equiv \epsilon_{kmn}s_{k}
\end {align}
so that
\begin {align} 
\label {al94}
\frac{ds_{k}}{d\tau}=0
\end {align}
again in agreement with eq. (\ref{al56}).
\par In the following section we describe the Hamiltonian formalism. The latter takes simpler form if we first reformulate slightly the Lagrangian approach. To this end assume we have solved explicitly the constraints (\ref{al1}) and expressed $\Lambda^{\mu}_{\phantom {\mu}\nu}$ in terms of six independent parameters $\phi^{a}$, $a=1,...,b$. Let us define \cite{b35}
\begin {align} 
\label {al95}
a_{a}^{\phantom{a}\mu\nu}\equiv \Lambda_{\beta}^{\phantom {\beta}\mu}\frac{\partial \Lambda^{\beta\nu}}{\partial \phi^{a}}\,\,\,\text{,}\,\,\, a_{a}^{\phantom{a}\mu\nu}=-a_{a}^{\phantom {a}\nu \mu}\quad \text{;}
\end {align}
$a_{a}^{\phantom {a}\mu \nu}$, considered as six by six matrix, is invertible because it relates two sets of six independent variables. Therefore, one can define its inverse by
\begin {align} 
\label {al96}
a_{a}^{\phantom{a}\mu\nu}b_{a}^{\phantom{a}\alpha \beta}=g^{\mu\alpha}g^{\nu\beta}-g^{\mu\beta} g^{\nu \alpha}
\end {align}
\begin {align} 
\label {al97}
a_{a}^{\phantom{a}\mu\nu}b_{b \mu \nu}=2\delta_{ab}
\end {align}
In terms of independent variables, $a^{\mu}$, $\phi^{a}$, our Lagrangian reads
\begin {align} 
\label {al98}
L=-m\Lambda_{\mu}^{\phantom {\mu}0}(\phi)\dot{a}^{\mu}+\frac{s}{2}\epsilon_{3kl}a_{a}^{\phantom{a}kl}(\phi)\dot{\phi}^{a}
\end {align}
and the equations of motion for $\phi^{a}$ take the form
\begin {align} 
\label {al99}
\frac{\partial L}{\partial \phi^{a}}-\frac{d}{d \tau} \Bigg (\frac{\partial L}{\partial \dot{\phi}^{a}}\Bigg)=0
\end {align}
or
\begin {align} 
\label {al100}
m \Lambda_{\mu \beta}a_{a}^{\phantom {a}\beta 0}\dot{a}^{\mu}+\frac{s}{2}\epsilon_{3kl}(a_{a \beta}^{\phantom {a \beta}l}a_{b}^{\phantom {b}\beta k}-a_{b \beta}^{\phantom {b \beta}l}a_{a}^{\phantom {a}\beta k})\dot{\phi}^{b}=0
\end {align}
Multiplying by $\Lambda_{\alpha \rho}\Lambda_{\mu \sigma}b_{a}^{\phantom {a}\rho \sigma}$ one obtains eq. (\ref{al71}).

\section{Hamiltonian formalism} 
\label{V}
\par Our starting point is the Lagrangian (\ref{al98}). The generalized momenta read
\begin {align} 
\label {al101}
P_{\mu}\equiv \frac{\partial L}{\partial \dot{a}^{\mu}}=-m \Lambda_{\mu}^{\phantom {\mu}0}
\end {align}
\begin {align} 
\label {al102}
\pi_{a}\equiv \frac{\partial L}{\partial \dot{\phi}^{a}}=\frac{s}{2}\epsilon_{3kl}a_{a}^{\phantom {a}kl}\quad \text {;}
\end {align}
note that the definition of $P_{\mu}$ differs by a sign from the one used in previous sections. 
\par Eqs. (\ref{al101}), (\ref{al102}) imply two kinds of primary constraints
\begin {align} 
\label {al103}
\widetilde{\Psi}_{1 \mu}\equiv P_{\mu}+m\Lambda_{\mu}^{\phantom {\mu}0}\approx 0
\end {align}
\begin {align} 
\label {al104}
\widetilde{\Psi}_{2a}\equiv \pi_{a}-\frac{s}{2}\epsilon_{3kl}a_{a}^{\phantom {a}kl}\approx 0
\end {align}
\par Due to the reparametrization invariance the Hamiltonian consists of the constraints only
\begin {align} 
\label {al105}
H\equiv \pi_{a}\dot{\phi}^{a}+p_{\mu}\dot{a}^{\mu}-L+u^{\mu}\widetilde{\Psi}_{1\mu}+v^{a}\widetilde{\Psi}_{2a}=u^{\mu}\widetilde{\Psi}_{1 \mu}+v^{a}\widetilde{\Psi}_{2a}
\end {align}
where $u^{\mu}$ and $v^{a}$ are the relevant Lagrange multipliers. The canonical Poisson brackets read
\begin {align} 
\label {al106}
\{\phi^{a},\pi_{b}\}=\delta ^{a}_{b}
\end {align}
\begin {align} 
\label {al107}
\{a^{\mu},P_{\nu}\}=\delta ^{\mu}_{\nu}
\end {align}
with all remaining brackets vanishing with the Hamiltonian being combination of primary constraints there are no secondary constraints. In fact, the consistency conditions read
\begin {align} 
\label {al108}
\dot{\widetilde{\Psi}}_{1 \mu}=\{\widetilde {\Psi}_{1 \mu},H\}\approx 0
\end {align}
\begin {align} 
\label {al109}
\dot{\widetilde{\Psi}}_{2a}=\{\widetilde {\Psi}_{2a},H\}\approx 0
\end {align}
or, explicitly, 
\begin {align} 
\label {al110}
\Lambda _{\mu \beta}v^{a}a_{a}^{\phantom {a}\beta 0}\approx 0
\end {align}
\begin {align} 
\label {al111}
mu^{\mu}\Lambda_{\mu \beta}a_{a}^{\phantom {a}\beta 0}+sv^{b}\epsilon_{3kl}a_{a}^{\phantom {a}k \beta}a_{b \beta}^{\phantom {b \beta}l}\approx 0
\end {align}
By multiplying eq. (\ref{al111}) by $b_{a}^{\phantom {a}\rho 0}$ and using eqs. (\ref{al96}) and (\ref{al110}) one obtains
\begin {align} 
\label {al112}
u^{\mu}\Lambda_{\mu k}\approx 0
\end {align}
which yields
\begin {align} 
\label {al113}
u^{\mu}=u\Lambda^{\mu}_{\phantom {\mu}0}
\end {align}
with $u$ being new Lagrange multiplier. On the other hand, by multiplying eq. (\ref{al111}) by $b_{a}^{\phantom {a}\rho k}$ and taking into account eq. (\ref{al110}) we get
\begin {align} 
\label {al114}
v^{a}a_{a}^{\phantom {a}\mu \nu}=0\quad \text {,}\quad (\mu \nu)\neq (12)(21)
\end {align}
\par The coefficients $a_{a}^{\phantom {a}\mu \nu}$, $\mu<\nu$ are independent so eqs. (\ref{al114}) provide five equations for six unknowns $v^{a}$. Therefore, the general solution reads
\begin {align} 
\label {al115}
v^{a}=vb_{a}^{\phantom {a}12}
\end {align}
with $v$ being again new Lagrange multiplier.
\par Eqs. (\ref{al105}), (\ref{al113}) and (\ref{al115}) allow us to write the final form of the Hamiltonian (up to irrelevant terms):
\begin {align} 
\label {al116}
H=u \Lambda^{\mu}_{\phantom {\mu}0}P_{\mu}+vb_{a}^{\phantom {a}12}\pi_{a}
\end {align}
The resulting canonical equations of motion read:
\begin {align} 
\label {al117}
\dot{a}^{\mu}=\{a^{\mu},H\}=u \Lambda^{\mu}_{\phantom{\mu}0}
\end {align}
\begin {align} 
\label {al118}
\dot{P}_{\mu}=\{P_{\mu},H\}=0
\end {align}
\begin {align} 
\label {al119}
\dot{\Lambda}^{\alpha}_{\phantom {\alpha}\beta}=\{\Lambda^{\alpha}_{\phantom {\alpha}\beta},H\}=v(\Lambda^{\alpha 1}\delta^{2}_{\beta}-\Lambda^{\alpha 2}\delta ^{1}_{\beta})
\end {align}
\begin {align} 
\label {al120}
\dot{\pi}_{a}=\{\pi_{a},H\}=-uP_{\mu}\Lambda^{\mu}_{\phantom {\mu}\alpha}a_{a}^{\phantom{a}\alpha 0}-v\pi_{b}\frac{\partial b_{b}^{\phantom {b}12}}{\partial \phi^{a}}
\end {align}
\par It is straightforward to check that the canonical equations (\ref{al121})-(\ref{al124}), together with the constraints
\begin {align} 
\label {al121}
P_{\mu}+m\Lambda_{\mu}^{\phantom {\mu}0}\approx 0
\end {align}
\begin {align} 
\label {al122}
\pi_{a}-\frac{s}{2}\epsilon_{3kl}a_{a}^{\phantom{a}kl}\approx 0
\end {align}
are equivalent to the Lagrange equations (\ref{al70}), (\ref{al71}). Only few remarks are in order here. It is immediate to see that the Lagrange equations follow from the Hamiltonian ones. To show the inverse implication note first that eq. (\ref{al74}) is equivalent to (\ref{al117}) upon the identification $u=\Lambda^{\mu}_{\phantom {\mu}0}\dot{a}_{\mu}$. Eq. (\ref{al70}) implies (\ref{al118}) provided the constraint (\ref{al121}) is taken into account. It remains to show that the eqs. (\ref{al119}), (\ref{al120}) result from Lagrange equations. Eqs. (\ref{al119}), when written out explicitly, read 
\begin {align} 
\label {al123}
\dot{\Lambda}^{\mu}_{\phantom {\mu}0}&=0\nonumber\\
\dot{\Lambda}^{\mu}_{\phantom {\mu}3}&=0\nonumber\\
\dot{\Lambda}^{\mu}_{\phantom {\mu}1}&=v\Lambda^{\mu}_{\phantom{\mu}2}\nonumber\\
\dot{\Lambda}^{\mu}_{\phantom {\mu}2}&=-v\Lambda^{\mu}_{\phantom{\mu}1}
\end {align}
First equation is already known to be valid. On the other hand, eq. (\ref{al75}) implies
\begin {align} 
\label {al124}
\Lambda^{\mu 1}\Lambda^{\nu 2}-\Lambda^{\mu 2}\Lambda^{\nu 1}\equiv d^{\mu \nu}=\text{const}
\end {align}
and, multiplying by $\Lambda_{\nu 2}$ or $\Lambda_{\nu 1}$:
\begin {align} 
\label {al125}
\Lambda^{\mu 1}=d^{\mu \nu}\Lambda_{\nu 2}
\end {align}
\begin {align} 
\label {al126}
\Lambda^{\mu 2}=-d^{\mu \nu}\Lambda_{\nu 1}
\end {align}
Differentiating (\ref{al125}) and (\ref{al126}) with respect to $\tau$ and using again (\ref{al124}) we find that two last equations (\ref{al123}) are fulfilled provided $v=-\Lambda^{\nu 1}\dot{\Lambda}_{\nu 2}$. Finally, we have
\begin {align} 
\label {al127}
\dot{\Lambda}_{\mu}^{\phantom{\mu}\nu}\Lambda^{\mu}_{\phantom {\mu}3}+\Lambda_{\mu}^{\phantom {\mu}\nu}\dot{\Lambda}^{\mu}_{\phantom {\mu}3}=0
\end {align}
For $\nu=0$ this gives $\Lambda_{\mu}^{\phantom{\mu}0}\dot{\Lambda}^{\mu}_{\phantom{\mu}3}=0$; also $\Lambda_{\mu}^{\phantom {\mu}3}\dot{\Lambda}^{\mu}_{\phantom{\mu}3}=0$. For $\nu = 1,2$ we use two last equations (\ref{al123}) to find $\Lambda_{\mu}^{\phantom {\mu}1,2}\dot{\Lambda}^{\mu}_{\phantom{\mu}3}=0$; so, finally $\Lambda_{\mu}^{\phantom {\mu}\alpha}\dot{\Lambda}^{\mu}_{\phantom {\mu}3}=0$ yielding second equation (\ref{al123}).
\par Let us now analyse the structure of constraints. Inspired by the form of general solution for Lagrange multipliers (\ref{al113}), (\ref{al115}) and the form of Hamiltonian (\ref{al116}) we define new equivalent set of constrains 

\begin {align} 
\label {al128}
\Psi_{1 \mu}\equiv \Lambda^{\nu}_{\phantom {\nu}\mu}\Psi_{1 \nu}=\Lambda^{\nu}_{\phantom {\nu}\mu}P_{\nu}+m \delta^{0}_{\mu}
\end {align}
\begin {align} 
\label {al129}
\Psi_{2}^{\phantom{2}(\mu \nu)}\equiv b_{a}^{\phantom {a}\mu \nu}\Psi_{2a}=b_{a}^{\phantom {a}\mu \nu}\pi_{a}-s\epsilon_{3kl}g^{\mu k}g^{\nu l}
\end {align}
Then the Hamiltonian takes a particularly simple form
\begin {align} 
\label {al130}
H=u\Psi_{10}+v\Psi_{2}^{\phantom {2}(12)}
\end {align}
It is straightforward to compute the Poisson brackets of new constraints:
\begin {align} 
\label {al131}
\{\Psi_{1 \mu},\Psi_{1 \nu}\}=0
\end {align}
\begin {align} 
\label {al132}
\{\Psi_{1\mu},\Psi_{2}^{\phantom {2}(\rho \sigma)}\}&=\delta_{\mu}^{\sigma}(\Psi _{1}^{\phantom {1}\rho}-m\delta_{0}^{\rho})-\delta_{\mu}^{\rho}(\Psi _{1}^{\phantom {1}\sigma}-m\delta_{0}^{\sigma})\nonumber\\
&\approx m(\delta_{\mu}^{\rho}\delta_{0}^{\sigma}-\delta_{\mu}^{\sigma}\delta_{0}^{\rho})
\end {align}
\begin {align} 
\label {al133}
\{\Psi_{2}^{\phantom {2}(\mu \nu)},\Psi_{2}^{\phantom{2}(\rho \sigma)}\}&=g^{\mu \rho}\Psi_{2}^{\phantom {2}(\nu \sigma)}-g^{\mu \sigma}\Psi_{2}^{\phantom {2}(\nu \rho)}+g^{\nu \sigma}\Psi_{2}^{\phantom {2}(\mu \rho)}-g^{\nu \rho}\Psi_{2}^{\phantom {2}(\mu \sigma)}\nonumber\\
&-s\epsilon_{3kl}(g^{\mu \rho}g^{\nu l}g^{\sigma k}-g^{\nu \rho}g^{\mu l}g^{\sigma k}-g^{\mu \sigma}g^{\nu l}g^{\rho k}+g^{\nu \sigma}g^{\mu l}g^{\rho k})\nonumber\\
&\approx -s \epsilon _{3kl}(g^{\mu \rho}g^{\nu l}g^{\sigma k}-g^{\nu \rho}g^{\mu l}g^{\sigma k}-g^{\mu \sigma}g^{\nu l}g^{\rho k}+g^{\nu \sigma}g^{\mu l}g^{\rho k})
\end {align}
All Poisson brackets become pure numbers on the constraint manifold. By inspecting eqs. (\ref{al131})-(\ref{al133}) we find that $\Psi_{10}$ and $\Psi_{2}^{\phantom{2}(12)}$ are first class constraints; the remaining ones are of second class. \\
$\Psi_{10}$ and $\Psi_{2}^{\phantom {2}(12)}$, being the first class primary constraints, generate gauge transformations. Consider first $\Psi_{10}$. One finds
\begin {align} 
\label {al134}
\delta a^{\mu}=\epsilon \{a^{\mu},\Psi_{10}\}=\epsilon\Lambda^{\mu}_{\phantom {\mu}0}
\end {align}
\begin {align} 
\label {al135}
\delta \Lambda^{\mu}_{\phantom {\mu}\nu}=\epsilon\{\Lambda^{\mu}_{\phantom {\mu}\nu},\Psi_{10}\}=0
\end {align}
On the other hand, $\Psi_{2}^{\phantom {2}(12)}$ generates the following transformations
\begin {align} 
\label {al136}
\delta a^{\mu}=\lambda\{a^{\mu},\Psi_{2}^{\phantom {2}(12)}\}=0
\end {align}
\begin {align} 
\label {al137}
\delta \Lambda^{\mu}_{\phantom {\mu}\nu}=\lambda\{\Lambda^{\mu}_{\phantom {\mu}\nu},\Psi_{2}^{\phantom {2}(12)}\}=-\lambda(\Lambda^{\mu}_{\phantom {\mu}1}\delta^{2}_{\nu}-\Lambda^{\mu}_{\phantom {\mu}2}\delta^{1}_{\nu})
\end {align}
By comparying the above equations with (\ref{al78})-(\ref{al80}) we conclude that the first class constraints generate the action of gauge group $G_{s}=SO(2)\times \mathbb{R}$. The reparametrization invariance is also a gauge symmetry. It is properly encoded in the formalism by the property that the Hamiltonian is a combination of first class constraints.
\par Let us note that the initial phase space is twentydimensional. We have ten constraints $\Psi_{1 \mu}$, $\Psi_{2}^{\phantom {2}(\mu \nu)}$ and two gauge degrees of freedom which leaves us with eightdimensional reduced phase space coinciding with the dimension of coadjoint orbit.\\
The eight constraints of second kind, $\Psi_{1i}$, $\Psi_{2}^{\phantom {2}(0i)}$, $\Psi_{2}^{\phantom {2}(13)}$, $\Psi_{2}^{\phantom {2}(23)}$, may be converted into strong equalities provided we replace the Poisson bracket by Dirac one. This can be done following the standard prescription yielding:
\begin {align} 
\label {al138}
\{A,B\}_{D}&=\{A,B\}-\frac{\Sigma}{M^{2}}\epsilon_{3ij}\{A,\Psi_{1i}\}\{\Psi_{1j},B\}\nonumber\\
&+\frac{1}{M}\Big(\{A,\Psi_{1i}\}\{\Psi_{2}^{\phantom {2}(0i)},B\}-\{A,\Psi_{2}^{\phantom {2}(0i)}\}\{\Psi_{1i},B\}\Big)\nonumber\\
&-\frac{1}{\Sigma}\epsilon_{3ij}\{A,\Psi_{2}^{\phantom {2}(i3)}\}\{\Psi_{2}^{\phantom {2}(j3)},B\}
\end {align}
where
\begin {align} 
\label {al139}
M \equiv \Psi_{10}-m
\end {align}
\begin {align} 
\label {al140}
\Sigma \equiv \Psi_{2}^{\phantom {2}(12)}+s
\end {align}
In what follows we assume the generic case $s\neq 0$; $s=0$ can be easily dealt with separately. Note that we can use the first class constraints, $M=-m$, $\Sigma=s$, only after all brackets have been already computed.
\par Using the general form (\ref{al138}) of Dirac bracket one computes the basic brackets:
\begin {align} 
\label {al141}
\{a^{\mu},a^{\nu}\}_{D}=\frac{\Sigma}{M^{2}}\epsilon_{3ij}\Lambda^{\mu i}\Lambda^{\nu j}
\end {align}
\begin {align} 
\label {al142}
\{a^{\mu},\Lambda^{\alpha}_{\phantom {\alpha}\beta}\}_{D}=\frac{1}{M}(g^{\mu \alpha}\delta^{0}_{\beta}-\Lambda^{\alpha}_{\phantom {\alpha}0}\Lambda^{\mu}_{\phantom {\mu}\beta})
\end {align}
\begin {align} 
\label {al143}
\{\Lambda^{\mu}_{\phantom {\mu}\nu},\Lambda^{\alpha}_{\phantom {\alpha}\beta}\}_{D}=\frac{1}{\Sigma}\epsilon_{3kl}(\Lambda^{\mu k}\delta^{3}_{\nu}-\Lambda^{\mu 3}\delta^{k}_{\nu})(\Lambda^{\alpha l}\delta^{3}_{\beta}-\Lambda^{\alpha 3}\delta^{l}_{\beta})
\end {align}
\par Eqs. (\ref{al141})-(\ref{al143}) provide the Poisson structure on the manifold of Poincare group. Additionally, we have the gauge symmetry generated by first class constraints which takes the following global form
\begin {align} 
\label {al144}
a'^{\mu}=a^{\mu}+\epsilon \Lambda^{\mu}_{\phantom {\mu}0}
\end {align}
\begin {align} 
\label {al145} 
\Lambda'^{\mu}_{\phantom {'\mu}1}=\Lambda^{\mu}_{\phantom {\mu}1}\cos \lambda +\Lambda^{\mu}_{\phantom {\mu}2} \sin \lambda
\end {align}
\begin {align} 
\label {al146} 
\Lambda'^{\mu}_{\phantom {'\mu}2}= -\Lambda^{\mu}_{\phantom {\mu}1}\sin \lambda +\Lambda^{\mu}_{\phantom {\mu}2} \cos \lambda
\end {align}
with $\epsilon$ and $\lambda$ being arbitrary parameters. Therefore, the manifold of gauge invariant elements is eightdimensional. We can easily show that it coincides with the coadjoint orbit described in Sec. III. To this end let us remind the definitions of $x_{i}$, $p_{i}$ and $s_{i}$. \textbf{Let us stress that these are definitions and not the constraints (in particular, note the minus sign as compared with the definition in this section).} We have 
\begin {align} 
\label {al147} 
p_{\mu}=m\Lambda_{\mu}^{\phantom {\mu}0}
\end {align}
\begin {align} 
\label {al148} 
s_{k}=sR_{k}^{\phantom {k}3}=s\Bigg(\Lambda_{k}^{\phantom {k}3}-\frac{\Lambda_{k0}\Lambda^{03}}{1+\Lambda^{0}_{\phantom {0}0}}\Bigg)
\end {align}
\begin {align} 
\label {al149} 
x_{k}=a_{k}-\frac{\Lambda_{k}^{\phantom {k}0}a_{0}}{\Lambda_{0}^{\phantom {0}0}}-\frac{\epsilon_{kln}s_{l}p_{n}}{m(m+p_{0})}
\end {align}
\par All these quantities are gauge invariant. Moreover, knowing $p_{\mu}$ and $s_{k}$ and using the normalization condition $\Lambda_{\mu}^{\phantom {\mu}3}\Lambda^{\mu 3}=-1$ one can compute $\Lambda^{\mu}_{\phantom {\mu}0}$ and $\Lambda^{\mu}_{\phantom {\mu}3}$. Now, the gauge transformations (\ref{al145}), (\ref{al146}) describe the rotations in the plane spanned by the fourvectors $\Lambda^{\mu}_{\phantom {\mu}1}$, $\Lambda^{\mu}_{\phantom {\mu}2}$ (they are fourvectors with respect to the left action of Lorentz group). Since $\Lambda^{\mu}_{\phantom {\mu}1}$, $\Lambda^{\mu}_{\phantom {\mu}2}$ are normalized and orthogonal no nontrivial invariant can be formed; only the orientation of the plane spanned by $\Lambda^{\mu}_{\phantom {\mu}1}$, $\Lambda^{\mu}_{\phantom {\mu}2}$ is a gauge invariant notion. However, the latter is determined by two orthogonal fourvectors $\Lambda^{\mu}_{\phantom {\mu}0}$ and $\Lambda^{\mu}_{\phantom {\mu}3}$. Finally, 
$x_{k}$ fixes the gauge invariant combinations of $a^{\mu}$. It is now straightforward to verify that the Dirac brackets of $x_{i}$, $s_{i}$, $p_{\mu}$, as defined by (\ref{al147})-(\ref{al149}), coincide with Poisson brackets (\ref{al49}),(\ref{al53}).
\par Finally, let us study in some detail the issue of Poincare invariance. We have found (cf. eqs. (\ref{al72}), (\ref{al73})) the conserved quantities following from Poincare symmetry. By virtue of eqs. (\ref{al43}), (\ref{al47}), (\ref{al48}) and (\ref{al147})-(\ref{al149}) we find 
\begin {align} 
\label {al150} 
\zeta_{\mu}=m \Lambda_{\mu}^{\phantom {\mu}0}
\end {align}
\begin {align} 
\label {al151} 
\zeta_{0i}=ma_{0}\Lambda_{i}^{\phantom {i}0}-ma_{i}\Lambda_{0}^{\phantom {0}0}+s\epsilon_{3kl}\Lambda_{0}^{\phantom {0}k}\Lambda_{i}^{\phantom {i}l}
\end {align}
\begin {align} 
\label {al152} 
\zeta_{ij}=ma_{i}\Lambda_{j}^{\phantom {j}0}-ma_{j}\Lambda_{i}^{\phantom {i}0}+s\epsilon_{3kl}\Lambda_{i}^{\phantom {i}k}\Lambda_{j}^{\phantom {j}l}
\end {align}
so that the generators of Poincare symmetry coincide with those found previously within the coadjoint orbit method. Obviously, they are gauge invariant. Let us take more close look at their action on Poincare group manifold. Consider first their action on translations. One has, by virtue of eq. (\ref{al142}), 
\begin {align} 
\label {al153} 
\delta a_{\mu}=\epsilon^{\nu}\{a_{\mu},\zeta_{\nu}\}=\epsilon^{\nu}\{a_{\mu},m \Lambda_{\nu}^{\phantom {\nu}0}\}=\epsilon^{\nu}(g_{\mu \nu}-\Lambda_{\mu}^{\phantom {\mu}0} \Lambda_{\nu}^{\phantom {\nu}0})
\end {align} 
or
\begin {align} 
\label {al154} 
\delta a_{\mu}=\epsilon_{\mu}-\Lambda_{\mu}^{\phantom {\mu}0}(\epsilon^{\nu}\Lambda_{\nu}^{\phantom {\nu}0})
\end {align} 
\par The above expression differs from standard form of translations, $\delta a_{\mu}=\epsilon_{\mu}$. However, one has to take into account that the part of $\delta a^{\mu}$ proportional to $\Lambda_{\mu}^{\phantom {\mu}0}$ is a pure gauge transformation (cf. eq. (\ref{al134})). The physically meaningful translation is obtained by subtracting the pure gauge part. This is the content of eqs. (\ref{al153}), (\ref{al154}). This is well known phenomenon; for example, in order to obtain the gauge invariant energy-momentum tensor in Maxwell theory one has to consider translations supplied with an appropriate gauge transformations.
\par Further, consider the Lorentz transformations of $a^{\mu}$. We find
\begin {align} 
\label {al155} 
\delta a_{\rho}=-\frac{1}{2}\omega^{\mu \nu}\{a_{\rho},\zeta_{\mu \nu}\}_{D}=\frac{1}{2}\omega^{\mu \nu}\big(g_{\mu \rho}a_{\nu}-g_{\nu \rho}a_{\mu}+\Lambda_{\rho}^{\phantom {\rho}0}(a_{\mu}\Lambda_{\nu}^{\phantom {\nu}0}-a_{\nu}\Lambda_{\mu}^{\phantom {\mu}0})\big)
\end {align} 
Again, although $a_{\mu}$ is a four-vector, one has to subtract pure gauge degree of freedom. In fact, for pure Lorentz transformation $\delta a_{\rho}=\omega^{\rho \nu}a_{\nu}$; then we should subtract the gauge part: $\delta a _{\rho}-\Lambda_{\rho}^{\phantom {\rho}0}(\Lambda_{\nu}^{\phantom {\nu}0} \delta a ^{\nu})$ which yields finally
\begin {align} 
\label {al156} 
\delta a_{\rho}=\frac{1}{2}\omega^{\mu \nu}\big(g_{\mu \rho}a_{\nu}-g_{\nu \rho}a_{\mu}+\Lambda_{\rho}^{\phantom {\rho}0}(a_{\mu}\Lambda_{\nu}^{\phantom {\nu}0}-a_{\nu}\Lambda_{\mu}^{\phantom {\mu}0})\big)
\end {align}
in full agreement with eq. (\ref{al155}).
\par Finally, consider the transformation properties of $\Lambda^{\alpha}_{\phantom {\alpha}\beta}$. One finds from eq. (\ref{al143})
\begin {align} 
\label {al157} 
\delta \Lambda^{\alpha}_{\phantom {\alpha}\beta}&=-\frac{1}{2}\omega^{\mu \nu}\{\Lambda^{\alpha}_{\phantom {\alpha}\beta},\zeta_{\mu \nu}\}_{D}=\nonumber\\
&=+\frac{1}{2}\omega^{\mu \nu}\big (\delta _{\mu}^{\alpha}\Lambda_{\nu \beta}-\delta _{\nu}^{\alpha}\Lambda_{\mu \beta}+\epsilon_{3kl}\Lambda_{\mu}^{\phantom {\mu}k}\Lambda_{\nu}^{\phantom {\nu}l}(\Lambda^{\alpha}_{\phantom {\alpha}1}\delta^{2}_{\beta}-\Lambda^{\alpha}_{\phantom {\alpha}2}\delta^{1}_{\beta})\big)
\end {align}
Also here we can understand the structure of eq. (\ref{al157}) in terms of gauge symmetry. Under the left action of Lorentz group each column $\Lambda^{\alpha}_{\phantom {\alpha}\beta}$, $\alpha=0,...,3$, $\beta$-fixed, transforms as a four-vector. Again, one has to subtract gauge part of variation; it is not difficult to see that the corrected transformation takes the form
\begin {align} 
\label {al158} 
\delta \Lambda^{\alpha}_{\phantom {\alpha}\beta}=\omega^{\alpha \nu}\Lambda_{\nu \beta}+\omega^{\mu \nu}\Lambda_{\mu}^{\phantom {\mu}1}\Lambda_{\nu}^{\phantom {\nu}2}(\Lambda^{\alpha}_{\phantom {\alpha}1}\delta^{2}_{\beta}-\Lambda^{\alpha}_{\phantom {\alpha}2}\delta^{1}_{\beta})
\end {align}
which agrees with eq. (\ref{al157}).
\section{General action functional} 
\label{VI}
\par The action functional (\ref{al65}) corresponds to a specific choice of the fixed spin vector $\vec{s}=(0,0,s)$. It may be generalized in a simple way by redefining the dynamical variables $\Lambda^{\mu}_{\phantom {\mu}\nu}$ multiplying them from the right by a fixed rotation $R^{\mu}_{\phantom {\mu}\nu}$:
\begin {align} 
\label {al159} 
\Lambda^{\mu}_{\phantom {\mu}\nu}\longrightarrow \Lambda'^{\mu}_{\phantom {'\mu}\nu}=\Lambda^{\mu}_{\phantom {\mu}\alpha}R^{\alpha}_{\phantom {\alpha}\nu}
\end {align}
Than the action functional takes the form
\begin {align} 
\label {al160} 
S=-\int \Bigg (m \Lambda_{\mu}^{\phantom {\mu}0}\dot{a}^{\mu}-\frac{1}{2}\underline{s}_{n}\epsilon_{nkl}\Lambda_{\mu}^{\phantom {\mu}k}\dot{\Lambda}^{\mu l}\Bigg)d\tau
\end {align}
with $\underline{\vec{s}}$ being an arbitrary vector with $\vert \underline{\vec{s}}\vert=s$.
\par Slightly different formulation can be also obtained as follows (cf. \cite{b19}, \cite{b33}). Let us define two fourvectors $k^{\mu}$, $l^{\mu}$ as follows
\begin {align} 
\label {al161} 
k^{\mu}\equiv \Lambda^{\mu 1}\nonumber\\
l^{\mu}\equiv \Lambda^{\mu 2}
\end {align}
They are obviously constrained by
\begin {align} 
\label {al162} 
k^{2}=-1\quad \text {,}\quad l^{2}=-1\quad \text {,}\quad k\cdot l=0
\end {align}
The gauge transformations (\ref{al145}), (\ref{al146}) read
\begin {align} 
\label {al163} 
k'^{\mu}&=k^{\mu}\cos \lambda +l^{\mu}\sin \lambda\nonumber\\
l'^{\mu}&=-k^{\mu}\sin \lambda +l^{\mu}\cos \lambda
\end {align}
As it has been already mentioned only the orientation of the plane spanned by $k^{\mu}$ and $l^{\mu}$ has a gauge invariant meaning. It may be characterized by choosing two fourvectors orthogonal both to $k^{\mu}$ and $l^{\mu}$. One of them is $p_{\mu}=m\Lambda_{\mu}^{\phantom {\mu}0}$ while the remaining one can be chosen as orthogonal also to $p_{\mu}$. Up to the sign and normalization we can take
\begin {align} 
\label {al164} 
n^{\mu}=\frac{1}{m}\epsilon^{\mu \nu \alpha \beta}p_{\nu}k_{\alpha}l_{\beta}
\end {align}
Then
\begin {align} 
\label {al165} 
n^{2}=-s^{2}\quad \text {,}\quad n\cdot p=n\cdot k =n\cdot l=0
\end {align}
and, actually
\begin {align} 
\label {al166} 
n^{\mu}=s \Lambda^{\mu}_{\phantom {\mu}3}=-s \Lambda^{\mu 3}
\end {align}
\par Now, the complete description of the relativistic spinning particle can be given in terms of the fourvector $n^{\mu}$. Indeed, taking into account the normalization $n^{2}=-1$ and orthogonality $n\cdot p=0$ conditions one concludes that $n^{\mu}$ carries two gauge invariant degrees of freedom. Together with three independent components of $p_{\mu}$ and three gauge invariant combinations of $a^{\mu'}s$, $a^{\mu}-\frac{p^{\mu}}{p^{0}}a^{0}$, we find that the phase space describing gauge invariant sector correctly reproduces the coadjoint orbit description.\\
Using eqs. (\ref{al147})-(\ref{al149}) and (\ref{al166}) it is easy to express $n^{\mu}$ in terms of $p_{i}$ and $s_{i}$
\begin {align} 
\label {al167} 
n_{0}&=-\frac{s_{k}p_{k}}{m}\nonumber\\
n_{i}&=-s_{i}-\frac{p_{i}p_{k}s_{k}}{m(m+p_{0})}
\end {align}
\par We conclude that $n_{\mu}$ is proportional to the Pauli-Lubanski fourvector
\begin {align} 
\label {al168} 
n_{\mu}=\frac{1}{m}w_{\mu}
\end {align}
\section{Quantum theory} 
\label{VII}
\par The classical dynamics described in the previous sections may be now canonically quantized. The most convenient starting point is provided by gauge invariant variables $x_{i}$, $p_{i}$ and $s_{i}$, obeying the Poisson algebra (\ref{al49})-(\ref{al53}). Canonical quantization procedure, $\{\,\,\, ,\,\,\}\rightarrow \frac{1}{i} [\,\,\, , \,\,]$, together with Stone-von Neumann theorem and representation theory of $SU(2)$ algebra, yields the following form of basic operators
\begin {align} 
\label {al169} 
\hat{p}_{i}=p_{i}\cdot \mathbb{1}
\end {align}
\begin {align} 
\label {al170} 
\hat{x}_{i}=\Bigg(+i\frac{\partial}{\partial p_{i}}-\frac{i p_{i}}{2p_{0}^{\phantom {0}2}}\Bigg)\cdot \mathbb{1}+iU(p)\frac{\partial U^{+}(p)}{\partial p_{i}}
\end {align}
\begin {align} 
\label {al171} 
\hat{s}_{i}=U(p)S_{i}U^{+}(p)\quad\text{;}
\end {align}
here $\{S_{i}\}$ are the matrices spanning some irreducible representation of the $SU(2)$ algebra, $\mathbb{1}$ is the corresponding unit matrix, $U(p)$ are arbitrary $p$-dependent unitary matrices belonging to this representation. The scalar product is defined as follows
\begin {align} 
\label {al172} 
(f,g)=\sum _{a}\int \frac{d^{3}\vec{p}}{2p_{0}}\,\overline{f_{a}(\vec{p})}g_{a}(\vec{p})
\end {align}
where $f_{a}(\vec{p})$ are wave functions taking values in the relevant representation space of $SU(2)$. Note that $\hat{x}_{i}$ is basically a covariant derivative corresponding to the trivial connection (this is in contrast with massless case where the monopole bundle emerges, (cf. \cite{b45}, \cite{b46}). In particular, choosing $U(p)\equiv \mathbb{1}$  one obtains
\begin {align} 
\label {al173} 
\hat{x}_{i}=i\frac{\partial}{\partial p_{i}}-\frac{ip_{i}}{2p_{0}^{\phantom {0}2}}
\end {align}
\begin {align} 
\label {al174} 
\hat{s}_{i}=S_{i}
\end {align}
Using eqs. (\ref{al47}), (\ref{al48}) it is now easy to find the form of Lorentz generators. There appears an ordering problem in defining boosts (\ref{al47}); it is, however, easily curable by making the simplest replacement $p_{0}x_{i}\rightarrow \frac{1}{2}(p_{0}x_{i}+x_{i}p_{0})$ which preserves the relevant communication rules. This yields, in our particular gauge (\ref{al173}), (\ref{al174})
\begin {align} 
\label {al175} 
\hat{M}_{0i}\equiv \hat{\zeta}_{0i}=ip_{0}\frac{\partial}{\partial p_{i}}+\frac{\epsilon_{ilk}S_{l}p_{k}}{m+p_{0}}
\end {align}
\begin {align} 
\label {al176} 
\hat{M}_{ij}\equiv \hat{\zeta}_{ij}=i\Bigg (p_{j}\frac{\partial}{\partial p_{i}}-p_{i}\frac{\partial}{\partial p_{j}}\Bigg)+\epsilon_{ijk}S_{k}
\end {align}
which coincides with the standard formulae (\ref{al45}).
\par It is also straightforward to check that the operator $\hat{x}_{i}$ is the Newton-Wigner coordinate operator \cite{b48}. Passing to the general form of basic operators, eqs. (\ref{al169})-(\ref{al171}), is equivalent to the replacement $M_{\mu \nu}\rightarrow U(p)M_{\mu \nu}U^{+}(p)$.

\section{Summary} 
\label{VIII}
\par We have presented fairly complete description of the classical dynamics of relativistic spinning particles based on the method of coadjoint orbits. The main technical tool was the explicit decomposition of arbitrary Lorentz matrix into the product of pure boost and rotation. A coadjoint orbit is isomorphic to some coset manifold. This allows to represent the Hamiltonian dynamics on such an orbit as constrained dynamics on group manifold (viewed as configuration space) exhibiting gauge symmetry related to the stability subgroup of some point on the orbit. We have performed complete analysis of the constrained dynamics on Poincare group showing its equivalence to the dynamics on coadjoint orbit. 
\par Due to the fact that all relevant dynamical variables are constructed explicitly the quantization procedure can be performed immediately leading to the explicit description of irreducible representations of Poincare group corresponding to massive particles. This yields the infinitesimal version of Wigner's procedure. In the Appendix we remind the standard Wigner algorithm and indicate its relation to the findings of Sec. VII.
\par In the forthcoming paper we will analyse, within the framework presented here, the model of relativistic spherical top proposed in the papers of Hanson and Regge \cite{b35}.

\appendix
\section*{Appendix}
\par We remind here the standard construction of unitary irreducible representations of Poincare group which correspond to positive mass $m$ and spin $s=0,\frac{1}{2},1,...$ . The space of states is spanned by the vectors $\vert \vec{p},\sigma \rangle$, $\vec{p}\in\mathbb{R}^{3}$, $\sigma = -s,...,s$ which form the complete orthonormal set,
\begin {align} 
\label {al177} 
\sum^{s}_{\sigma = -s}\int \frac{d^{3}\vec{p}}{2p_{0}}\vert \vec{p},\sigma \Cross\vec{p},\sigma\vert=\mathbb{1}
\end {align}
\begin {align} 
\label {al178} 
\langle\vec{p},\sigma\vert \vec{p}\,',\sigma'\rangle =2p_{0}\delta
^{(3)}(\vec{p}-\vec{p}\,')\delta_{\sigma \sigma'}
\end {align}
The relevant wave functions are given by
\begin {align} 
\label {al179} 
f_{\sigma} (\vec{p})\equiv \langle\vec{p},\sigma\vert  f\rangle \quad\text {;}
\end {align}
then
\begin {align} 
\label {al180} 
(f,g)=\sum^{s}_{\sigma=-s}\int\frac{d^{3}\vec{p}}{2p_{0}}\overline{f_{\sigma}(\vec{p})}\, g_{\sigma} (\vec{p})
\end {align}
Let $k=(m,\vec{0})$ and let $L(p)$ be a standard boost, i.e.
\begin {align} 
\label {al181} 
p^{\mu}=L^{\mu}_{\phantom {\mu}\nu}(p)k^{\nu}=mL^{\mu}_{\phantom {\mu}0}(\vec{p})
\end {align}
We define the states $\vert\vec{p},\sigma\rangle$ by
\begin {align} 
\label {al182} 
\vert\vec{p},\sigma\rangle\equiv U\big(L(p)\big)\vert k,\sigma\rangle
\end {align}
Due to
\begin {align} 
\label {al183} 
U(\Lambda,a)=U(a)U(\Lambda)
\end {align}
it is sufficient to define separately the action of translations and Lorentz group. The former reads
\begin {align} 
\label {al184} 
U(a)\vert\vec{p},\sigma\rangle=e^{ia^{\mu}p_{\mu}}\vert\vec{p},\sigma\rangle
\end {align}
while the definition (\ref{al182}) leads to
\begin {align} 
\label {al185} 
U(\Lambda)\vert\vec{p},\sigma\rangle=\sum^{s}_{\sigma'=-s}D_{\sigma'\sigma}\big(R(p,\Lambda)\big)\vert\Lambda p,\sigma'\rangle
\end {align}
with $D_{\sigma'\sigma}(...)$ being spin $s$ representation of $SU(2)$ while
\begin {align} 
\label {al186} 
R(p,\Lambda)=L^{-1}(\Lambda p)\Lambda L(p)\in SO(3)
\end {align}
is the so-called Wigner rotation. The corresponding transformation of the wave functions takes the form
\begin {align} 
\label {al187} 
\big (U(\Lambda)f\big)_{\sigma}(\vec{p})=\sum^{s}_{\sigma'=-s}D_{\sigma \sigma'}\big(\tilde{R}(p,\Lambda)\big)f_{\sigma'}(\Lambda^{-1}p)
\end {align}
with
\begin {align} 
\label {al188} 
\tilde{R}(p,\Lambda)=R^{-1}(p,\Lambda^{-1})
\end {align}
There is a freedom in the choice of basic vectors. To see this let us note that eq. (\ref{al181}) does not define $L(p)$ uniquely. Let $R(\vec{p}): \mathbb{R}^{3}\rightarrow SO(3)$ be an arbitrary (smooth) function; then $L(p)R(\vec{p})$ also obeys eq. (\ref{al181}). Therefore, one can define a new basis
\begin {align} 
\label {al189} 
\vert\vec{p},\sigma\rangle^{R}\equiv U\big(L(p)\big)U\big(R(\vec{p})\big)\vert k,\sigma\rangle=\sum^{s}_{\sigma'=-s}D_{\sigma' \sigma}\big(R(p)\big)\vert\vec{p},\sigma'\rangle
\end {align}
\par Then the modified transformation rule takes the form
\begin {align} 
\label {al190} 
U(\Lambda)\vert\vec{p},\sigma\rangle^{R}=\sum^{s}_{\sigma'=-s}D_{\sigma' \sigma}\big(R^{-1}(\Lambda p)R(p,\Lambda)R(p)\big)\vert\Lambda p,\sigma'\rangle^{R}
\end {align}
\par The counterpart of eq. (\ref{al189}) for the wave function reads
\begin {align} 
\label {al191} 
f^{R}_{\sigma}(\vec{p})=D_{\sigma\sigma'}\big(R^{-1}(p)\big)f_{\sigma'}(p)
\end {align}
leading to the modified transformation law
\begin {align} 
\label {al192} 
\big(U(\Lambda)f^{R}\big)_{\sigma}(\vec{p})=\sum^{s}_{\sigma'=-s}D_{\sigma\sigma'}\big(R^{-1}(p)\tilde{R}(p,\Lambda)R(\Lambda^{-1}p)\big)f^{R}_{\sigma'}(\Lambda^{-1}p)
\end {align}
\par Once the global transformation laws are defined one can ask about their infinitesimal form. To this end one puts $\Lambda^{\mu}_{\phantom {\mu}\nu}=\delta^{\mu}_{\nu}+\omega^{\mu}_{\phantom {\mu}\nu}$ and expands the transformation formulae to the first order in $\omega$. In this way the structure described in Sec. VII emerges.
\\
\\
{\bf Acknowledgements}
\\
The research has been supported by the grant 2016/23/B/ST2/00727 of National Science Center, Poland.

\end {document}